\documentclass[useAMS,twocolumn]{emulateapj}
\usepackage[latin2]{inputenc}
\pdfoutput=1

\newcommand{\br}{\mathbf{r}}
\newcommand{\hmpc}{$~h^{-1}$Mpc}
\newcommand{\muK}{$\mathrm{\mu K}$}
\newcommand{\LCDM}{$\Lambda$CDM}

\newcommand{\tisw}{T_{\small \rm ISW}}
\newcommand{\tiswi}{T_{{\rm ISW},i}}
\newcommand{\tiswj}{T_{{\rm ISW},j}}
\newcommand{\nside}{n_{\rm side}}

\newcommand{\tcmb}{T_{\small \rm CMB}}
\newcommand{\vbar}{\overline{V}}
\newcommand{\xbar}{\overline{x}}

\def\figzhist{figs/redshift_hist_lat.pdf}
\def\figdust{figs/dustz.pdf}
\def\figgal{figs/gal.pdf}
\def\figisw{figs/pot.pdf}
\def\figcmb{figs/cmb.pdf}
\def\figslicecl{figs/zcuts_cls.pdf}
\def\fignoise{figs/noisemodel.pdf}
\def\figslices{figs/noisemodel_zwidth.pdf}
\def\figcf{figs/crossisw.pdf}
\def\figdiff{figs/clean_cf.pdf}
\def\figprof{figs/profiles.pdf}
\def\figsup{figs/sup.pdf}

\def\eg{e.g.\ }

\begin{document}

\title{A Map of the Integrated Sachs-Wolfe Signal from Luminous Red Galaxies}

\author{Benjamin R. Granett\altaffilmark{1}, Mark C. Neyrinck\altaffilmark{1,2} and Istv\'an Szapudi\altaffilmark{1,3,4}}
\altaffiltext{1}{Institute for Astronomy, University of Hawaii, 2680 Woodlawn Drive, Honolulu HI 96822, USA}
\altaffiltext{2}{Department of Physics and Astronomy, The Johns Hopkins University, 3701 San Martin Drive, Baltimore, MD 21218, USA}
\altaffiltext{3}{Institute for Advanced Study, Collegium Budapest, Szentháromság u.\ 2., Budapest, H-1014, Hungary}
\altaffiltext{4}{Eötvös Loránd University, Dept.\ of Atomic Physics, 1117 Pázmány Péter sétány 1/A}

\begin{abstract}
We construct a map of the time derivative of the gravitational
potential traced by SDSS Luminous Red Galaxies.  The potential decays
on large scales due to cosmic acceleration, leaving an imprint on
cosmic microwave background (CMB) radiation through the integrated
Sachs-Wolfe (ISW) effect.  With a template fit, we directly measure
this signature on the CMB at a 2-$\sigma$ confidence level.  The
measurement is consistent with the cross-correlation statistic,
strengthening the claim that dark energy is indeed the cause of the
correlation.  This new approach potentially simplifies the
cosmological interpretation.  Our constructed linear ISW map shows no
evidence for degree-scale cold and hot spots associated with supervoid
and supercluster structures.  This suggests that the linear ISW effect
in a concordance \LCDM~cosmology is insufficient to explain the strong
CMB imprints from these structures that we previously reported.

\end{abstract}


\section{Introduction}
Large-scale structures in the low-redshift Universe leave a mark
on the cosmic microwave background (CMB) radiation through the
late-time integrated Sachs-Wolfe (ISW) effect \citep{sachswolfe}.  In
\LCDM, the expansion of the Universe accelerates at late times,
causing gravitational potentials to decay.  The effect gives an energy
boost to photons traveling through massive structures and degrades the
energy of photons crossing under-dense voids.  In a flat universe, the
effect occurs only in the presence of dark energy \citep{crittenden}.
The evolution of the ISW signal provides a constraint on the dark
energy equation of state, and is a unique probe of cosmology,
independent from studies using supernovae as standard candles, the
baryon acoustic scale, galaxy cluster counts, or weak lensing
\citep{corasaniti05,pogo05,Dent08}.

The linear ISW signal on the CMB is difficult to measure because the
temperature fluctuations from the ISW effect are about an order of
magnitude smaller than those of the primary CMB.  However, the effect
can be measured statistically by correlating the structure at low
redshift with the CMB temperature.  This has been carried out with
many galaxy samples with redshift $z<2$, including the APM survey
\citep{fosalba04}, 2MASS \citep{afshordi04,rassat07}, SDSS Luminous
Red Galaxies \citep{scranton03,fosalba03,pad05}, SDSS Main Sample
\citep{cabre06}, SDSS quasars \citep{gian06}, NVSS radio sources
\citep{Boughn04,Nolta04,rac2008} and the X-ray background
\citep{Boughn04}.  The correlation has been studied using wavelet
analyses as well \citep{McEwen06,Piet06,mcewen08}.  Most recently, the
combined analysis of multiple surveys gives a detection significance
$\gtrsim 4\sigma$ \citep{Ho08,gian08}.

On non-linear scales, gravitational potentials evolve through
structure formation processes, giving rise to the Rees-Sciama (RS)
effect \citep{RS68,sakai08}.  The RS effect dominates over the linear
ISW signal at small angular scales (multipoles $>1000$) on the CMB
\citep{cooray02b,cai08}.  Other secondary anisotropies on the CMB
include the Sunyaev-Zeldovich (SZ) effect induced by inverse-Compton
scattering of CMB photons by hot gas in massive clusters
\citep{SZ}. Gravitational lensing by foreground structures also
induces a temperature anisotropy, although both of these effects are
restricted to arcminute scales on the CMB \citep{cooray02}.  On high
redshift galaxy samples, the interpretation of degree scale matter-CMB
correlations can be complicated by magnification bias, which boosts
the ISW signal \citep{Loverde07}, as well as large-scale correlations
arising at reionization \citep{gian07}.

In this paper, we construct an ISW map from a sample of luminous red
galaxies (LRGs), assuming linear growth of density fluctuations, and
using a parameter-free Voronoi tessellation technique to add the
potential from each galaxy directly.  We measure the ISW signal on the
CMB using the constructed map as a template.  Typically, the ISW
measurement is degraded by cosmic variance on the CMB temperature as
well as variance in the local large-scale structure.  Our approach
utilizes information about the observed density field traced by the LRG
survey to remove the effects of local cosmic variance, potentially
improving the detection.  We compare this technique with the
cross-correlation statistic.  For the LRG sample, the two measurements
are consistent, although the error analysis is simplified considerably
by the template fit approach because a Monte-Carlo analysis is not
required.

Template fits (or matched filter analyses) are commonly employed in
studying Galactic foreground emission on the CMB \citep{oliveira99},
and have been used in the context of the SZ effect \citep{Hansen}.
The application to ISW has been proposed by a number of groups,
including \citet{hernandez08}, \citet{frommert08}, and
\citet{barreiro08}.  \citeauthor{hernandez08} and
\citeauthor{frommert08} find that by removing cosmic variance from the
measurement, the ISW detection significance can improve by 10\% over
the correlation function measurement, a conclusion also supported by
\citet{cabre07}.

A map of the foreground anisotropies is also of interest in itself.
Large structures, especially at low redshift, can potentially produce
significant anisotropies on the CMB \citep{maturi07,sakai08}.  These
might explain CMB anomalies including the alignment of low multipoles
\citep{inouesilk07}, and the 5\degr~Cold Spot \citep{rudnick07}.
\cite{supervoids} found a 4-$\sigma$-significant signature
corresponding to supervoids and superclusters in in the {\it Wilkinson
  Microwave Anisotropy Probe} (WMAP) maps, which hints at physics
outside of \LCDM~\citep[\eg][]{hunt08}. We will further investigate
this signal again using template fitting techniques.

Unless noted, we employ joint WMAP5, supernovae and BAO $\Lambda$CDM
parameters: $\Omega_{c}=0.233$, $\Omega_{b}=0.0462$, $h=0.701$ with
$\sigma_8=0.817$ \citep{komatsu}.

\section{Data}
\subsection{Luminous Red Galaxies}
\begin{figure}
    \begin{center}
      \includegraphics[scale=1.0]{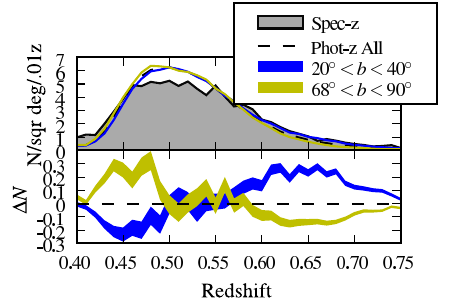}
    \end{center}  
    \caption{The redshift distribution of the LRG catalog binned with
      $\Delta z=.01$ is plotted. We constructed the photometric
      catalog from SDSS DR6 using the same selection criteria as the
      2SLAQ spectroscopic LRG survey.  The renormalized spectroscopic
      redshift distribution is plotted in the top frame (shaded).
      Also plotted are the photometric redshift distributions of a
      high Galactic latitude sample with $b>68\degr$ and a low
      latitude cut with $20\degr<b<40\degr$.  The two samples contain
      equal numbers of galaxies.  The difference in number density
      between the full and subsamples are plotted in the bottom frame.
      The shaded range represents Poisson errors.  We find that the
      low latitude sample has a systematically higher median redshift,
      but is within the redshift error
      distribution. \label{fig:redshift}}
\end{figure}

We base our study on the Sloan Digital Sky Survey (SDSS) \citep{SDSS}
Luminous Red Galaxy (LRG) sample.  The photometric galaxy catalog
traces large-scale structure to $z=0.8$ over 7500 square degrees of
contiguous sky. LRGs are elliptical galaxies in massive galaxy
clusters representing large dark-matter halos \citep{Blake}, and are
thought to be physically similar objects across their redshift range
\citep{Eisenstein2001,Wake2006}.  This makes them excellent, albeit
sparse, tracers of the cosmic matter distribution on scales $\gtrsim
10$ Mpc.  Our sample is designed to match the spectroscopic targets in
the 2SLAQ LRG survey \citep{Cannon06}.  In particular, we apply the
stringent $i<19.8$ and $d_{perp}>0.55$ cuts.  Within the DR6 imaging
area about the North Galactic Pole, the catalog includes 746,962
objects; for our fiducial sample, though, we use the 400,000 closest
to the median in redshift.  Photometric redshifts are obtained from
\citet{Oyaizu08}; we use the results from the CC2 algorithm.  The
median photometric error of the sample is $\sigma_z=.039$.  The
redshift distribution is plotted in Fig. \ref{fig:redshift}.  It
extends from $0.45<z<0.75$ with median $z=0.52$.  A map of the
projected density is shown in Fig. \ref{fig:maps}.

\begin{figure}
    \begin{center}
      \includegraphics{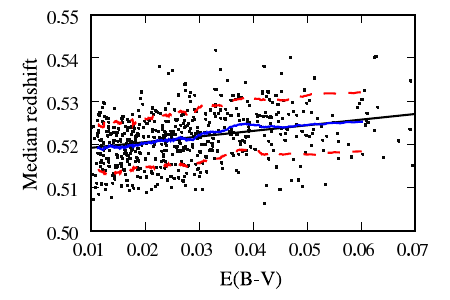}
    \end{center}  
    \caption{We attribute the shift of the photometric redshift
      distribution with Galactic latitude to residual errors in the
      extinction correction.  Shown is the median redshift in
      12\degr~bins (Healpix pixels at $\nside$-16) plotted against the
      Galactic extinction, E(B-V)\citep{Schlegel}.  Over-plotted are
      the mean and one standard deviation limits, as well as a linear
      fit.  The correlation coefficient is $r=0.38$.  This trend is an
      order of magnitude below the size of the redshift errors, but
      systematically degrades the 3D density construction; see \S4.3.
    \label{fig:gradient}}
\end{figure}

We find a systematic trend in the photometric redshift distribution
with Galactic latitude; see Fig. \ref{fig:redshift}.  At low latitude,
the distribution shifts toward higher redshift.  Although the shift in
the median redshift is smaller than the typical redshift error, it
produces a strong gradient in the galaxy density at high and low
redshift where the selection function is steep, which contaminates the
3D density reconstruction.  We attribute this to residual errors in
the Galactic extinction correction.  The trend of median redshift with
extinction, E(B-V), is shown in Fig. \ref{fig:gradient}.  Due to this
issue, the useable redshift range is limited to the peak of the
distribution, as we discuss in \S4.3.

\subsection{Microwave maps}
We use WMAP five-year maps, quoting results from the the Q, V and W
frequency band maps, in which the individual differencing assembly
maps have been co-added.  To estimate Galactic foreground
contamination, we use the foreground-reduced Q, V and W maps, the
Markov-chain Monte Carlo (MCMC) derived temperature map and the
Internal Linear Combination (ILC) map, all of which are described by
\citet{gold08}.  The Q, V and W foreground reduced maps are
full-resolution maps with best-fit foreground templates subtracted,
preserving the noise properties of the original maps.  The MCMC map
was produced using a Monte Carlo joint fit of the CMB temperature,
polarization and foregrounds.  The map is smoothed at
1\degr~full-width half-max resolution.  The ILC map is
foreground-cleaned using a minimum-variance fit of the foreground
templates.  It is smoothed to 1\degr~resolution.  We present the
results from the ILC map even though its noise properties are not
characterized.  To each map we apply the KQ75 galactic foreground and
point source mask.  We carry out all analyses using the
Healpix\footnote{\url{http://healpix.jpl.nasa.gov}} pixelization
scheme at a resolution of 55\arcmin/pixel ($\nside=64$)
\citep{healpix}, at finer resolution than the $>1$\degr~ scales where
the linear ISW effect is thought to operate.  The MCMC map is shown in
Fig. \ref{fig:maps}.

\section{ISW map construction}
The ISW-induced temperature is an integral over the time-derivative of
the gravitational potential $\Phi$.  In terms of the conformal time
$\tau$, the temperature fluctuation $\tisw$ along a line of sight, is
\citep{sachswolfe}
\begin{equation}
\frac{\Delta \tisw}{\tcmb} = -\frac{2}{c^2} \int d \tau \frac{d \Phi}{d \tau}.
\label{deltat-lineofsight}
\end{equation}

In linear theory, $\delta$ evolves with the growth factor $D_1(z)$ as
$\delta(\vec{x}, z)=\frac{D_1(z)}{D_1(z=0)}\delta(\vec{x}, z=0)$.  In
a flat, matter dominated universe $D_1 = 1/(1+z)$, but in \LCDM, it
deviates from this at low redshift due to accelerated expansion.  We
obtain the time derivative of the potential assuming linear theory,
i.e.

\begin{equation}
\frac{d\Phi ({\bf x})}{d\tau} = \frac{\Phi({\bf x})}{(1+z)}\frac{d}{d\tau}\left[(1+z)D_1(z)\right].
\end{equation}

We model the density field in the survey using a Voronoi tessellation
\citep[e.g.][]{okabe,vdws}.  Each galaxy occupies a polyhedral Voronoi
cell of points closer to that galaxy than to any other.  The volume
$V_i$ of that cell gives a natural estimate of the galaxy's
overdensity, $\delta_i=\vbar(z_i)/V_i - 1$, where $\vbar(z)$ is the
average volume of a galaxy at redshift $z$.

We compute the potential as a direct sum over these polyhedral cells
of volume $V_i$ and overdensity $\delta_i$, concentrated at the
comoving positions $\br_i$ of their galaxies.  In the Newtonian limit,
the potential is
\begin{eqnarray}
  \Phi({\bf r}) & = & -\frac{G\rho_{{\rm cr},0}\Omega_m}{c^2}(1+z_{\rm med})\sum_i \frac{\delta_i V_i}
    {|{\br} - {\br_i}|}\\
 & = & -\frac{G\rho_{{\rm cr},0}\Omega_m}{c^2}(1+z_{\rm med})\sum_i \frac{\vbar(z_i) - V_i}
    {|{\br} - {\br_i}|}.
    \label{potential}
\end{eqnarray}
Here, $\rho_{{\rm cr},0}$ is the critical density at $z=0$.  In
evaluating this expression, we put the whole survey at the median
redshift of the sample, $z_{\rm med}=0.52$, multiplying by a factor
$\frac{(z_{\rm med}+1)D_1(z_{\rm med})}{(z+1)D_1(z)}$.  After
computing the potential, we then divide $\Phi$ by this factor to get
the proper amplitude.  Explicitly, the observed density is the
true density convolved with the photometric redshift error distribution,
and attenuated by the top-hat survey window function:
$\delta = (\hat{\delta}\star f_z)\times W({\bf r})$.

Our method is relatively slow computationally, since it involves a sum
over all galaxies for each point where the potential is evaluated, and
a Voronoi tessellation.  For our fiducial galaxy sample, the
computation takes a few hours using several CPU's, while a fast
fourier transform method could take less than a minute on one CPU.
However, our method easily allows the true (curved) geometry of the
LRG sample to be included, and integrated through to $z=0$.  Also,
the contribution to the potential from each galaxy is included in a
natural way using the parameter-free, highly adaptive Voronoi method.

We added a buffer of galaxies around the survey, and filled survey
holes with galaxies sampled at the mean density at each redshift, as
in \citet{supervoids}.  Hole galaxies (comprising $\sim$ 1/300 of the
galaxies) were included in the potential sum, but galaxies neighboring
any buffer galaxy (i.e.\ affected by the edge) were not.  To get the
mean galaxy volume function $\vbar(z)$, we spline-interpolated between
galaxy volumes averaged in bins of logarithmic width $d\log_{10} r =
1/200$.

\begin{figure*}
      \includegraphics[scale=.95]{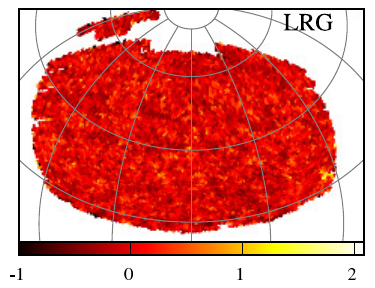}
      \includegraphics[scale=.95]{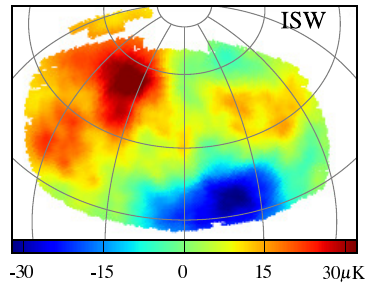}
      \includegraphics[scale=.95]{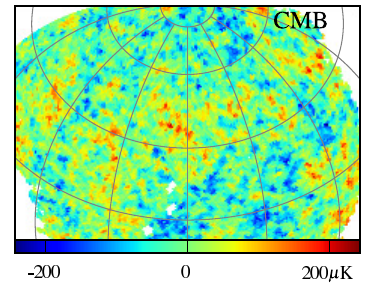}
    \caption{The maps used in our analysis.  The left frame is the LRG
      galaxy overdensity within the SDSS DR6 survey footprint.  At
      center is the reconstructed ISW signal in \muK.  Right is the
      foreground-cleaned CMB temperature from the WMAP5 MCMC analysis
      with the galactic foreground and point source mask applied.  The
      maps are in celestial coordinates with 55\arcmin/pixel.  The
      grid spacing is 30\degr\ with RA=180\degr\ at the center, and RA
      increasing to the left.
      \label{fig:maps}}
\end{figure*}

To get the predicted ISW map, we integrated the signal
(Eq.\ (\ref{deltat-lineofsight})) through lines of sight of Healpix
pixels.  We integrated from $z=0$ to 1.5 times the farthest distance
of a galaxy in the survey, sampling the potential at 10 (a sufficient
number, according to convergence tests) equally spaced points in each
of three regions: in front of, within, and behind the sample.  In
spots, the nonzero potential behind and in front of the survey generated
$\sim$50\% of the signal.  To lessen the noise from galaxies happening
to lie too close to a point $\br$ where the potential was evaluated,
we softened the potential by setting the distance to a galaxy to
max$(|{\br} - {\br_i}|,~ 10$\hmpc).  The resulting map is shown in
Fig. \ref{fig:maps}.

A constant multiplicative bias factor $b_g$ is thought to describe the
ratio of the galaxy to matter overdensity adequately on the large
scales of the linear ISW effect: $\delta_{DM}=(1/b_g) \delta_g$.  In
the case of the LRG sample, $b_g\sim 2.2$ ($\sigma_8=0.82$)
\citep{Blake}.  We expect the true ISW fluctuations $\tisw$ to be
lower than our construction by a factor of $b_g$: $\tisw^{\rm (true)}
= (1/b_g) \tisw^{\rm (LRG)}$.  This dependence on the bias allows for
$b_g$ to be fit directly, independent of the power spectrum amplitude.

\section{Statistical significance}
\subsection{Amplitude fit}
To judge the reality of the reconstructed ISW signal, we measure its
contribution to the microwave sky with a template fit approach.  The
observed microwave temperature is the sum of the primary CMB
anisotropy, galactic and extragalactic foreground contributions and
the ISW signal, $\tisw^{\rm (LRG)}/b_g$.  The ISW anisotropy arises
from the structure sampled by LRG galaxies corrected for the galaxy
bias.  We add the predicted ISW template with a scaling parameter
$\lambda$, to be fit:
\begin{equation}
T = T_{\rm prim} + T_{\rm fg} + \frac{\lambda}{b_g} \tisw.
\end{equation}
Fixing the ISW and foreground anisotropies, the likelihood function,
in terms of the pixel CMB covariance matrix, $C_{ij}=\langle T_{{\rm
  prim},i} T_{{\rm prim},j}\rangle$, is,
\begin{eqnarray}
\ln \mathcal{L} = -\frac{1}{2}\sum_{ij} (T_i - T_{{\rm fg},i} - \frac{\lambda}{b_g} \tiswi)C_{ij}^{-1} \nonumber & &\\
\times (T_j - T_{{\rm fg},j} - \frac{\lambda}{b_g} \tiswj)& &,
\end{eqnarray}
and maximizing this over $\lambda$ gives,
\begin{equation}
 \hat{\lambda} = b_g \frac{\sum_{ij}(T_i-T_{{\rm fg},i}) C_{ij}^{-1} \tiswj}{\sum_{ij}\tiswi C_{ij}^{-1} \tiswj}.
\label{eq:fit}
\end{equation}
 The variance of this estimator is,
\begin{equation}
\sigma^2 = b_g^2 \frac{1}{\sum_{ij} \tiswi C_{ij}^{-1} \tiswj}.
\label{eq:error}
\end{equation}

This approach can be extended naturally to multiple correlated
datasets through a joint fit of many template maps.  The estimator
is non-linear in the the ISW template.  Thus, shot noise, arising in
the galaxy survey, leads to a biased estimate of the amplitude.
Perturbing the ISW template by $\tisw'=\tisw + \epsilon$, leads to an
underestimate of the amplitude.  Given shot noise variance,
$\sigma^2/N_{gal}$, that scales with the number of galaxies, the
expected amplitude takes the form,

\begin{equation}
\langle \lambda' \rangle = \langle \lambda \rangle \frac{1}{1 + \sigma^2/N_{gal}}.
\label{eq:noise}
\end{equation}

In the case of the ISW effect, the covariance matrix is dominated by
the primary CMB.  We estimate the covariance matrix using the best-fit
angular power spectrum from WMAP5 convolved with the pixel window
function and instrument beam: $C_{ij}=\langle x_i x_j \rangle = \sum_l
\frac{2l + 1}{4\pi}C_l W_l^2 B_l^2 P_l(\cos\theta_{ij})$, where $W_l$
is the pixel window function, and $B_l$ is the instrument beam.  Pixel
noise is added according to the noise properties of the map
considered.  In the case of the ILC map, pixel noise is neglected
because it is not characterized.

A correction to the covariance must be added to address the fact that
we estimate the mean of the distribution from the sample itself.  The
covariance is $C_{ij}'=\langle (x_i-\xbar) (x_j-\xbar) \rangle$ where
$\xbar$ is the mean pixel value within the survey:
$\xbar=\frac{1}{N_{pix}}\sum_i x_i$.  Expanding, this becomes
$C_{ij}' = \langle x_i x_j \rangle - \frac{1}{N}\sum_\alpha \langle
x_i x_\alpha\rangle - \frac{1}{N}\sum_\alpha \langle x_\alpha x_j
\rangle + \frac{1}{N^2}\sum_\alpha \sum_\beta \langle x_\alpha x_\beta
\rangle.$ Although this correction is $<5\%$ for the SDSS survey area,
neglecting it leads to an overestimate in the significance of the ISW
signal.  

Finally, the computed covariance matrix is ill-conditioned, leading to
an unstable inversion.  This is generally the case for a map covering
only a fraction of the sky.  We find a usable pseudo-inverse by
computing the singular value decomposition of the matrix and zeroing
the noisiest modes.  The eigenvalue limit was chosen such that a
stable inversion was produced for all CMB maps we considered.

\subsection{Results}
The calculation is carried out over the intersection of the galaxy
survey footprint and CMB foreground mask, in total 9298 pixels.

We tested ISW maps generated from four redshift cuts listed in Table
\ref{table:zcuts} fit to the MCMC CMB map.  The error was determined
with Eq. (\ref{eq:error}).  We find that the signal degrades in the
wider redshift ranges in samples III and IV.  The maximum signal was
found in sample II, which includes $\sim2/3$ of the galaxies about the
median redshift, spanning $0.48<z<0.58$.  Including all galaxies
reduces the significance by $0.6\sigma$.  We attribute this to
contamination from a large-scale gradient in the galaxy density
arising from a systematic shift in the redshift distribution discussed
in \S2.1.  This error is not accounted for in our analysis and we
limit ourselves to the sample II redshift range for our results.  The
differences between the ranges that we test are comparable to the size
of the photometric redshift error, so adjusting the cut should not
significantly bring in new structures affecting the intrinsic signal.

\begin{figure}
    \begin{center}
      \includegraphics{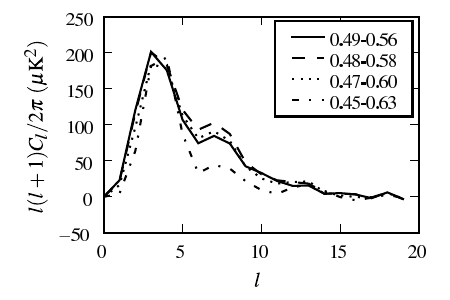}
    \end{center}  
    \caption{Plotted is a comparison of the power spectra of ISW maps
      constructed with different redshift samples, see Table
      \ref{table:zcuts}.  The maps were multiplied by the best-fit
      amplitude, $\lambda$.  This factor properly normalizes the maps,
      which otherwise show a variation in amplitude.  The signal
      primarily arises from low multipoles, $l<10$.
    \label{fig:zrangecls}}
\end{figure}

The power spectra of the ISW maps are shown in
Fig. \ref{fig:zrangecls}.  The maps are multiplied by the best-fit
$\lambda$ value from Table \ref{table:zcuts}.  This normalization
properly scales the spectra, aligning the peak at multipole $l=4$.
This scaling property with $\lambda$ suggests that narrowing the
redshift range acts as a smoothing kernel, washing out fluctuations in
the potential.  We further investigate this effect with simulations in
\S4.3.

We calculated the best-fit ISW amplitude for the Q, V and W maps, as
well as for the various foreground reduced maps using sample II.  The
results are listed in Table \ref{table:templatefit}, and are all
consistent with a 2$\sigma$ signal.

There is little evidence for foreground contamination.  Although there
is a 0.2$\sigma$ difference between the V and W maps, the amplitudes
found from the foreground-reduced maps agree with the raw co-added
maps.  This suggests that the slight color dependence is not due to
foreground contamination, but may arise from other differences such as
the map beams and noise properties.  Furthermore, the ILC and MCMC
maps agree with each other, and match the V filter.  This finding is
consistent with previous ISW studies, including \citet{Ho08} who place
an upper limit of 0.3$\sigma$ on the effect of foreground
contamination on the ISW detection significance.

\citet{Li2009} found a large-scale correlation in WMAP data between
the number of observation passes made over a region of sky and the
derived temperature which could affect the ISW measurement.  To check
for this, we computed the template fit between the number of
observations of each pixel and the temperature of the Q map at
$\nside=64$ resolution.  Over the SDSS area, we find no correlation on
the mean pixel temperature with $\lambda=(6\pm40)\times10^{-6} \mu
{\rm K} /N_{obs}$.  It is still possible that there is an effect from
the disparity in observation number between the plus and minus
antennae as reported by \citet{Li2009} which will require further
investigation.

\subsection{Interpretation}
The constructed potential is affected by the survey window function,
photometric redshift errors and shot noise.  We account for these
effects with the aid of simulations.  We generate mock LRG catalogs
from Gaussian large-scale structure simulations extending to $z=2$
with 40-Mpc cells and model the SDSS LRG sample as a slab cut out from
this volume.  We used a fast fourier transform to compute the
potential in this rectangular geometry, enabling the results of
hundreds of simulations to be averaged together.  The template fit in
this case measures the contribution of the survey slab's ISW signal to
the full ISW signal integrated from $z=0$ to 2.  We model redshift
errors by convolving the density field with a Gaussian kernel with
$\sigma_{\rm z}=0.04$ along the radial direction and introduce shot
noise by sampling the field according to the survey selection
function.

\begin{figure}
    \begin{center}
      \includegraphics{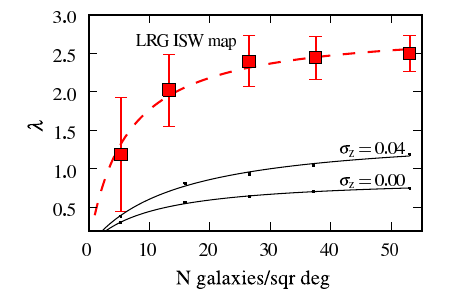}
    \end{center}  
    \caption{Shot noise in the galaxy survey leads to an
      underestimation of the amplitude.  Plotted is $\lambda$ measured
      from a galaxy field with the given projected number density.
      The expected value of $\lambda$ is 1.  The square markers are
      mean measurements of the ISW amplitude made on the under-sampled
      LRG catalog, which, when fully sampled, contains 53
      galaxies/square degree.  The error bars represent the sampling
      error on a single realization. The curves are fits to
      Eq.\ (\ref{eq:noise}).  The solid curves represent mean results
      from mock LRG catalogs drawn from Gaussian simulations, with
      photometric redshift errors modeled by a Gaussian kernel with
      dispersion $\sigma_{\rm z}$.
    \label{fig:noisetest}}
\end{figure}

Fig. \ref{fig:noisetest} shows the effect of shot noise in the
simulations and on the LRG measurement.  We under-sampled the LRG
catalog to create ISW maps with increased shot noise and fit the trend
with Eq.\ (\ref{eq:noise}).  Although the amplitude measured from LRGs
is higher than from simulations, the trends are similar.  Shot noise
biases the measurement low; in Fig. \ref{fig:noisetest}, the
asymptotic limit to $\lambda$ from the LRGs is 2.8, to be compared
with the measured result from the full map, 2.51, in Table
\ref{table:templatefit}.

\begin{figure}
    \begin{center}
      \includegraphics{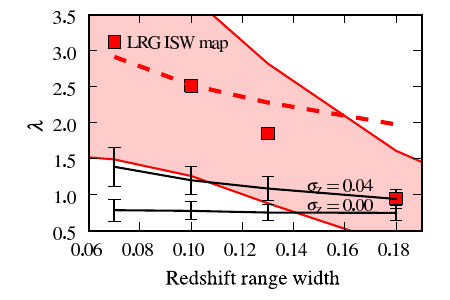}
    \end{center}
    \caption{We simulate the effect of redshift cuts on the derived
      ISW amplitude $\lambda$ (solid lines).  The square markers show
      our measurement on LRGs, as in Table \ref{table:zcuts}.  We find
      that introducing redshift errors boost the measured amplitude by
      a factor of 2.  The effect of shot noise is a downward bias in
      $\lambda$, making the expected value less than 1 when
      $\sigma_{\rm z}=0$.  The dashed curve is the simulation result
      multiplied by 2.1, demonstrating that our measurement is a
      factor of 2 larger than expected in \LCDM, although the
      measurement deviates from the simulation trend in the wider
      redshift ranges due to systematic redshift errors (see text).
      The error bars on the simulations represent sampling error on a
      single realization. \label{fig:zrange} }
\end{figure}

The redshift error convolution and survey window function act to bias
$\lambda$ high.  This can be understood due to the smoothing effect
redshift errors have on the reconstructed potential.
Fig. \ref{fig:zrange} shows the results of the simulations using the
four tested redshift cuts.  The measured trend in $\lambda$ with
survey width does not agree with the simulation results, but we
attribute the deviation in the wider redshift samples to the
large-scale gradient in the galaxy density in the tail of the redshift
distribution (discussed in \S2.1), and so, we do not consider these
points for the cosmological interpretation.

After modeling both shot noise and redshift error effects, we make a
simple correction to our measurements by normalizing $\lambda$ by the
results measured from simulations.  In sample II, $\lambda$ is
corrected by a factor of 1.2.  The corrected values are listed in
Tables \ref{table:zcuts} and \ref{table:templatefit}.  Our measurement
is a factor of $\sim2.1$ above the \LCDM~simulation results, within
the 1-$\sigma$ range of cosmic variance.  This result is in
concordance with the higher-than-expected ISW amplitude found by
\citet{Ho08}, \citet{gian08} and others, and is consistent with our
own correlation function measurements presented in \S5.

\begin{table}
\begin{center}
\caption{Dependence on redshift cuts\label{table:zcuts}}
\begin{tabular}{lccccc}
\tableline
    & redshift  & N        & Amplitude   & Amplitude &  \\
    &   range   & $/10^3$  & uncorrected & corrected & $\sigma$ \\
\tableline
I   & 0.49-0.56 & 300 & $3.12 \pm 1.63$ & 2.2 & 1.9 \\
II  & 0.48-0.58 & 400 & $2.51 \pm 1.25$ & 2.1 & 2.0 \\
III & 0.47-0.60 & 500 & $1.85 \pm 0.97$ & 1.7 & 1.9 \\
IV  & 0.45-0.63 & 600 & $0.95 \pm 0.66$ & 1.0 & 1.4 \\
\tableline
\end{tabular}
\end{center}
\end{table}

\begin{table}
\begin{center}
\caption{Amplitude fits\label{table:templatefit}}
\begin{tabular}{lccc}
\tableline
     & Amplitude   & Amplitude &  \\
Map  & uncorrected & corrected & $\sigma$ \\
\tableline
Q Coadd  & $2.40 \pm 1.19$ & 2.0 & 2.0\\
V Coadd  & $2.51 \pm 1.19$ & 2.1 & 2.1\\
W Coadd  & $2.13 \pm 1.17$ & 1.8 & 1.8\\
\tableline
Q FG reduced& $2.33 \pm 1.19$ & 1.9 & 2.0\\
V FG reduced& $2.51 \pm 1.19$ & 2.1 & 2.1\\
W FG reduced& $2.20 \pm 1.17$ & 1.8 & 1.9\\
\tableline
MCMC& $2.51 \pm 1.25$ & 2.1 & 2.0\\
ILC & $2.51 \pm 1.25$ & 2.1 & 2.0\\
\tableline
\end{tabular}
\end{center}
\end{table}

\section{Cross-correlation amplitude}
\subsection{Galaxy - CMB temperature correlation}
The standard measure of the ISW effect is the galaxy overdensity-CMB
temperature cross-correlation.  The measurement has been made on the
SDSS LRG dataset at the $2-2.5\sigma$ level \citep{Ho08,gian08}.

The expected angular power spectrum in the flat-sky limit, see \eg
 \citet{afshordi04}, is an integral over the line-of-sight distance
$r$,
\begin{eqnarray}
C_l^{\delta T} &=& \tcmb\frac{3 H_0^2 \Omega_m b_g}{c^2} \frac{1}{(l+1/2)^2} \nonumber \\
&&\times \int dr\ r^2 n(r) \frac{d(1+z)D_1(z)}{dr} P\left(\frac{l+1/2}{r}\right),
\end{eqnarray}
where, $n(r)=\frac{dN(r)}{dz dV}$ is the galaxy selection function
normalized by $\int r^2 \frac{dN}{dz dV}dr$, $D_1$ is the growth
factor and $P\left(z, k=\frac{l+1/2}{r}\right)$ is the matter power spectrum at
redshift $z$.  The correlation function in real space is,
\begin{equation}
w^{\delta T}(\theta)=\sum_l \frac{2l + 1}{4\pi}C^{\delta T}_l P_l(cos \theta),
\end{equation}
where $P_l$ are the Legendre polynomials.

We measure the goodness-of-fit against a fiducial model scaled to the
data.  A one parameter fit is sufficient because variations in
cosmology affect the ISW amplitude, but have little effect on the
shape of the spectrum \citep{pad05}.  We apply the same template fit
estimator (Eq. (\ref{eq:fit})) in the fashion of \citet{Ho08}: given an
observed data vector, $X$, and model template, $Y$, we find the best
fit amplitude, $\lambda$, for the model $X=\lambda Y$.  The maximum
likelihood estimate of the scale amplitude $\lambda$ is
$\hat{\lambda}=\frac{X C^{-1} Y}{Y C^{-1} Y}$, with variance
$\sigma^2=(Y C^{-1} Y)^{-1}$, where $C=\langle X_i X_j\rangle$ is the
covariance matrix.  In the case of the angular power spectrum,
$C=\langle C_l C_{l'}\rangle$.  Given a fixed cosmological model, the
free scaling is a constraint on the product of the galaxy bias and
normalization of the power spectrum, $b_g \sigma_8^2$.

We use SpICE \citep{szapudi01} to estimate the $C_l$ spectrum for
multipoles $l=4-192$, the limits of which are set by the resolution of
the map and the extent of the survey.  
The measured angular power spectrum is binned into 15 logarithmic band
powers.  

We consider only CMB variance in the error analysis.  Neglecting the
cosmic variance in the galaxy field leads to an underestimate of the
errors by 10\% \citep{cabre07}.  We estimate the covariance matrix in a
Monte-Carlo procedure with 2000 realizations of the CMB generated
according to the best fit WMAP5 angular power spectrum.  We consider
here the correlation with the MCMC CMB map, which we model using an
instrumental beam with 1\degr~full-width half-max.

The cross-correlation results for the MCMC map are plotted in
Fig. \ref{fig:cf}.  The best fit amplitude is
$b_g\sigma_8^2=2.82\pm1.35$, a $2.1\sigma$ measurement.  This is
consistent with previous results using the LRG
dataset\citep{Ho08,gian08}.  We find that the signal is a factor of 1.9
greater than the \LCDM~prediction with $b_g$=2.2 and $\sigma_8=0.817$.

\subsection{ISW-cleaned CMB map}
We now ask whether the constructed ISW map contains the signal found
in the galaxy-CMB cross-correlation.  We produce a cleaned CMB map by
subtracting the constructed ISW template. The template is scaled by
the best fit amplitude from Table \ref{table:templatefit}.  The
cross-correlation of the cleaned CMB map with the LRG map is shown in
Fig. \ref{fig:diff}.  We find that the subtraction effectively nulls
the correlation signal.  This agreement strengthens the claim that the
correlation is of the form expected from the ISW effect.

\subsection{ISW - CMB temperature correlation}
\begin{figure*}
    \begin{center}
      \includegraphics{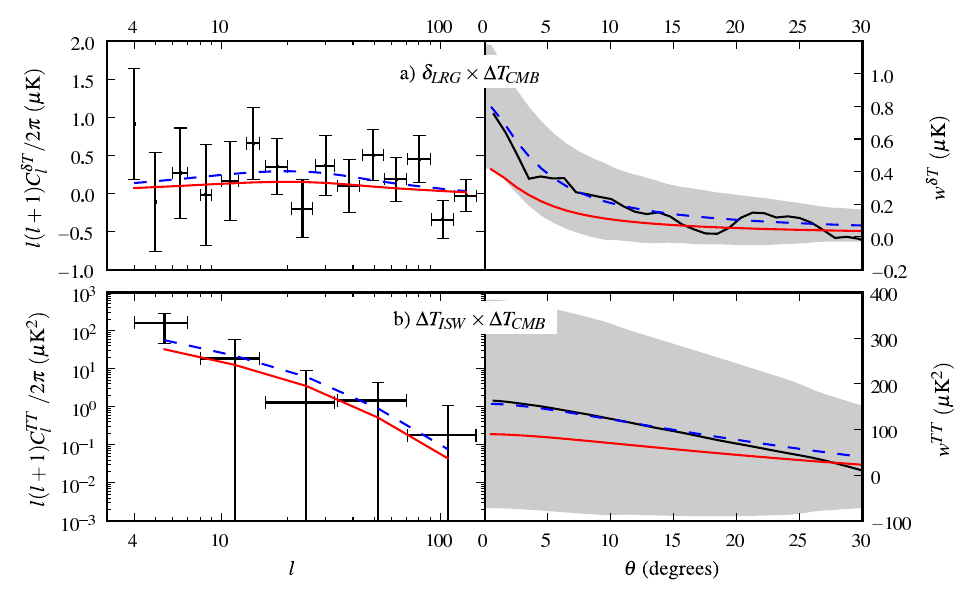}
    \end{center}  
    \caption{The ISW signal may be measured with the CMB temperature -
      galaxy cross-correlation function, plotted in panel a, top.  The
      left frame shows the cross power spectrum in harmonic space,
      which corresponds to the real-space correlation function on the
      right.  Over-plotted are \LCDM~ template ISW models: the dashed
      blue line is the template scaled by the best fit amplitude and
      the solid red line is the \LCDM~ prediction with $\sigma_8=0.82$
      and $b=2.2$.  Panel b shows the correlation of the constructed
      ISW map with the CMB, which measures the ISW auto-power
      spectrum.
    \label{fig:cf}}
\end{figure*}

\begin{figure}
    \begin{center}
      \includegraphics{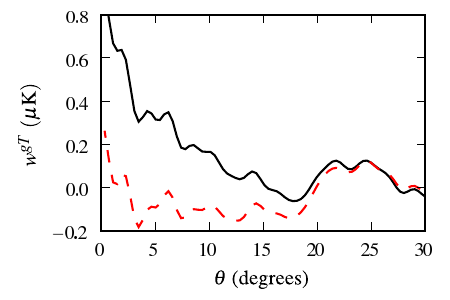}
    \end{center}  
    \caption{Subtracting the constructed ISW anisotropy from the CMB
      map nulls the correlation.  The solid curve shows the original
      galaxy-CMB cross correlation.  After subtracting the ISW signal,
      it is reduced to the dashed curve.  The measurement error is
the same as in Fig. \ref{fig:cf}.
    \label{fig:diff}}
\end{figure}

The ISW power spectrum gives a further measure with which to characterize
our ISW construction.  The cross-correlation between the reconstructed
ISW map and the CMB temperature measures the ISW auto-power spectrum, and,
by construction it is boosted by one power of the galaxy bias: $b_g
C_l^{TT}$.  In the flat-sky limit the expected spectrum, \eg
\citet{cooray02}, is,
\begin{eqnarray}
C_l^{TT} &=& \tcmb^2 \left(\frac{3 H_0^2 \Omega_m}{c^2}\right)^2 \frac{1}{(l+1/2)^4} \nonumber \\
&&\times \int dr\ r^2 \left(\frac{d(1+z)D_1(z)}{dr}\right)^2 P\left
(\frac{l+1/2}{r}\right).
\label{iswcl}
\end{eqnarray}
The ISW power is dominated by structures at low redshift which are not
captured by our map.  To account for this, we limit the integral in
Eq. (\ref{iswcl}) from $z=0.48-0.58$, the approximate redshift range of
the survey.  We carry out a fit to this model, again using Monte-Carlo
realizations of the CMB to estimate the covariance.  The measurement
is plotted in Fig. \ref{fig:cf}.  The best-fit amplitude is a factor of
$ 1.74\pm1.46$ above \LCDM.  Though the detection is marginal, it is
consistent with the amplitudes measured above.

\section{Supervoids and Superclusters}

Recently we detected the imprints of voids and clusters on the CMB and
attributed the finding to the linear ISW effect \citep{supervoids}.
The same SDSS LRG dataset at $z=0.5$ discussed here was used in that
work, although with minor differences in the selection criteria and
photometric redshifts.  Thus, with the ISW signal reconstruction,
we can further investigate the result and place limits on the role of
linear ISW in the measurement.  We also reassess the significance of the
measurement with a template fit on the CMB.

\begin{figure}
    \begin{center}
      \includegraphics[scale=1]{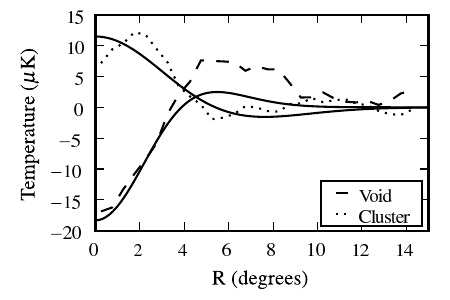}
      \includegraphics[scale=1.]{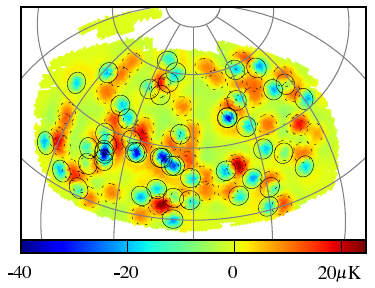}
    \end{center}  
    \caption{\label{fig:sup}Left: the measured void and cluster
      temperature profiles on the CMB from \citep{supervoids}.  Right:
      we use the mean profiles to generate a map of the voids and
      clusters on the sky.  Voids are blue and marked by a solid
      circle; clusters are red and marked by a dashed circle.  We
      assess the significance of these temperature fluctuations with a
      template fit, confirming the original finding.}
\end{figure}

The 50 voids and 50 clusters presented by \citet{supervoids} are
strongly correlated with the CMB.  The clusters, on average, fall on
hot spots on the CMB, while the voids are cold.  The mean temperature
profiles on the CMB are plotted in Fig. \ref{fig:sup}.  We re-examine
the significance of the correlation with the template-fitting analysis
used above, using a template constructed with the mean profiles of the
voids and clusters.  A compensated model was chosen to fit the
profiles to ensure that the mean of the map is zero.  For simplicity
we use a ``Mexican hat'' Laplacian of a Gaussian for the functional,
fitting for the amplitude and width.  The template fit confirms the
$4\sigma$ significance of the measurement found previously: we find
$\lambda=1.2\pm0.25$.  This high signal is not surprising because the
void and cluster profiles were measured on the CMB, although it does
confirm the $4\sigma$ peculiarity of these sites on the CMB.

If this correlation is due to the linear ISW effect, we would expect
the signal to be contained in the constructed ISW map presented here.
However, we find that the mean temperature of the clusters and voids
on the ISW map is not significant.  The temperature difference between
clusters and voids is $T^{\rm cluster}-T^{\rm voids}=0.08\pm0.1 \mu
K$, and stacking the clusters and voids separately does give weak hot
and cold spots.  We measure the temperature in a 4\degr~ compensated
aperture.  The error bar was measured through a Monte-Carlo procedure
using random locations within the survey.  This result contradicts the
suggestion that the signal is due to the linear ISW effect.

\section{Conclusions}
We implement a template fit measurement of the ISW signature imprinted
on the CMB from structures traced by SDSS luminous red galaxies at
$z\sim 0.5$ with photometric redshifts.  The approach uses the density
information in the galaxy survey to remove the effects of local cosmic
variance from the ISW measurement.  The detection significance is
2$\sigma$, which is consistent with the cross-correlation statistic.

Our measured amplitude confirms the cross-correlation finding, that
the signal is 1$\sigma$ higher than the \LCDM~prediction.  The
evidence suggests that this amplitude arises from variance on the CMB
\citep{Ho08}, or alternative ISW physics, although we cannot exclude
cosmic variance in the galaxy sample.

In principle, the template fit simplifies the error analysis and
cosmological interpretation, because a Monte-Carlo procedure is not
required to assess the role of cosmic variance in the galaxy sample,
and multiple datasets can be combined naturally.  On the other hand,
the measurement is influenced by a number of systematic effects that
do hinder the way towards precision cosmology.  First, contamination
by foreground emission must be subtracted from the CMB and the
accuracy of the subtraction ultimately limits the ISW detection.  This
is also true of cross-correlation function studies.  Second, shot
noise in the galaxy survey biases the estimator, but in a largely
correctable fashion.  In the case of the LRG survey, we estimate that
the effect is $<10\%$.  Third, the photometric redshift errors
severely degrade the reconstructed density and potential 3D maps,
leading to a biased estimate of the template amplitude.  In
simulations of the LRG catalog, the template signal was reduced by a
factor of 2.  We expect that this bias can be mitigated by using
cosmological density reconstruction techniques that take the redshift
errors as a Bayesian prior \citep[\eg][]{Ensslin}, and the effect can
be corrected with simulations.

The ISW template fit provides a unique constraint on cosmological
parameters.  In particular, because the ISW reconstruction is made
from a biased tracer, the template fit constrains the galaxy bias
independently of the the power spectrum amplitude, $\sigma_8$.
Precision results from the ISW signal will become feasible with new
all-sky survey projects, such as Pan-STARRS \citep{panstarrs}.

In this context, we further investigated our measurement of imprints
of $\sim 100$ Mpc supervoids and superclusters on the CMB
\citep{supervoids}.  We confirm that the signal is present in WMAP
data at $4\sigma$ confidence with a template fit.  However, we find
that the signal has much lower amplitude in the linear ISW map
constructed from LRGs than in the WMAP data. 

\acknowledgments We thank Adrian Pope for useful discussions and for
sharing his SDSS expertise. Some of the results were derived with
CosmoPy (\url{http://www.ifa.hawaii.edu/cosmopy}) and Healpix
\citep{healpix}.  We acknowledge the use of the LAMBDA archive
(\url{http://lambda.gsfc.nasa.gov}).  We are grateful for support from
NASA grant NNG06GE71G and NSF grant AMS04-0434413. Additional
appreciated support is provided to MCN by a grant from the
W.\ M.\ Keck Foundation at the Johns Hopkins University, and to IS by
the Pol\'anyi Program of the Hungarian National Office for Research
and Technology (NKTH).

Funding for the SDSS and SDSS-II has been provided by the Alfred
P. Sloan Foundation, the Participating Institutions, the National
Science Foundation, the U.S. Department of Energy, the National
Aeronautics and Space Administration, the Japanese Monbukagakusho, the
Max Planck Society, and the Higher Education Funding Council for
England. The SDSS Web Site is \url{http://www.sdss.org/}.

The SDSS is managed by the Astrophysical Research Consortium for the
Participating Institutions. The Participating Institutions are the
American Museum of Natural History, Astrophysical Institute Potsdam,
University of Basel, University of Cambridge, Case Western Reserve
University, University of Chicago, Drexel University, Fermilab, the
Institute for Advanced Study, the Japan Participation Group, Johns
Hopkins University, the Joint Institute for Nuclear Astrophysics, the
Kavli Institute for Particle Astrophysics and Cosmology, the Korean
Scientist Group, the Chinese Academy of Sciences (LAMOST), Los Alamos
National Laboratory, the Max-Planck-Institute for Astronomy (MPIA),
the Max-Planck-Institute for Astrophysics (MPA), New Mexico State
University, Ohio State University, University of Pittsburgh,
University of Portsmouth, Princeton University, the United States
Naval Observatory, and the University of Washington.

\bibliographystyle{hapj}
\bibliography{refs}

\end{document}